\documentclass[prd,aps,a4paper,twocolumn]{revtex4}

\usepackage{graphicx,psfrag}
\usepackage{mathrsfs}
\newcommand\p{\partial}
\renewcommand\l{\lambda}
\newcommand\s{\sigma}

\begin{document}

%%%%%%%%%%%%%%%%%%%%%%%%%%%%%%%%%%%%%%%%%%%%%%%%%%%%%%%%%%%%%

\title{Generic behaviour of nonlinear sound waves near the surface of
  a star: smooth solutions}

\author{Carsten Gundlach and Colin Please}

\affiliation{School of Mathematics, University of Southampton,
Southampton, SO17 1BJ, UK}

\date{18 January 2009, revised version 10 April 2009}

\begin{abstract}
We are interested in the generic behaviour of nonlinear sound waves as
they approach the surface of a star, here assumed to have the
polytropic equation of state $P=K\rho^\Gamma$. Restricting to
spherical symmetry, and considering only the region near the surface, we
generalise the methods of Carrier and Greenspan (1958) for the shallow
water equations on a sloping beach to this problem. We give a
semi-quantitative criterion for a shock to form near the surface
during the evolution of generic initial data with support away from
the surface. We show that in smooth solutions the velocity and the
square of the sound speed remain regular functions of Eulerian radius
at the surface.
\end{abstract}

\maketitle

%\tableofcontents

%%%%%%%%%%%%%%%%%%%%%%%%%%%%%%%%%%%%%%%%%%%%%%%%%%%%%%%%%%%%%

\section{Introduction}

%%%%%%%%%%%%%%%%%%%%%%%%%%%%%%%%%%%%%%%%%%%%%%%%%%%%%%%%%%%%%

In numerical simulations of neutron stars in general relativity, the
matter is often modelled as a perfect fluid. The simplest
equation of state usually considered is the ideal gas equation of
state
\begin{equation}
P=(\Gamma-1)e\rho,
\end{equation}
where $P$ is the pressure, $\rho$ the rest mass density and $e$ the
internal energy per rest mass. The polytropic index $\Gamma\equiv
1+1/n>1$ is a constant. 

If the entropy per rest mass is everywhere the same, the ideal gas
equation of state reduces to the polytropic equation of state
\begin{equation}
\label{polytrope}
P=K\rho^\Gamma,
\end{equation}
where $K$ is another constant depending on the entropy per rest
mass. If the initial data are isentropic, the solution remains
isentropic until a shock forms. For the polytropic equation of state
with $n>0$, spherically symmetric self-gravitating solutions with a
regular centre (stars) have a surface, characterised by $P=\rho=0$ at
finite radius $r=r_*$, where $\rho\sim (r_*-r)^n$ near the surface.

Standard numerical methods for evolving stars fail at the surface
because division by zero density occurs and the speed of sound goes to
zero. For smooth solutions in spherical symmetry, this can be avoided
by using Lagrangian coordinates, but in 3-dimensional (3D) simulations
with high-resolution shock capturing (HRSC) methods, the standard
practice is to match the star to a thin ``atmosphere'', which is then
artificially kept from accreting onto it. This method is likely to
give qualitatively wrong results, as the wave structure of the Riemann
problem that underlies HRSC methods is different if the right state is
vacuum.

The failure of the numerical methods is related to the physical fact that the
perfect fluid approximation must break down at the surface. This
approximation includes the assumption that small fluid elements are in
thermal equilibrium on dynamical timescales, but as the density goes
to zero, the thermal timescale diverges while the fluid dynamical
timescales are still determined by waves in the interior and remain
finite. In reality, some kind of plasma physics approximation
applies. 

The premise of this paper is that a mathematically correct numerical
implementation of the perfect fluid assumption is more correct than
the use of an unphysical atmosphere, which at best introduces
physically unmotivated approximations and at worst does not even have
a continuum limit. In this paper we provide two mathematical results
that should be useful in achieving this goal. We begin here with
smooth solutions and leave shocks for later work. 

Our preliminary question is whether smooth initial data representing
an outgoing wave with compact support form a shock as the wave
approaches the surface. That a shock forms is suggested by the fact
that the sound speed goes to zero at the surface with $c_s \sim
\sqrt{r_*-r}$ (independently of the polytropic index $n$), so that any
outgoing wave steepens. Sperhake \cite{Sperhake} has investigated this
numerically in general relativity in spherical symmetry and concludes
that small amplitude waves do not shock but large amplitude waves
do. In the Newtonian case in spherical symmetry this had already been
proved by Pelinovsky and Petrukhin \cite{PP}.  We improve on this
result by deriving a semiquantitative criterion for a sound wave to
remain regular as it approaches the surface.

Our main question is what kinematic boundary conditions can be used
in a numerical simulation to represent the free boundary at the
surface of the star. This has been addressed in general relativity by
Sperhake \cite{Sperhake} for nonlinear spherical perturbations, using
Lagrangian coordinates, and by Passamonti \cite{Passamonti} for linear
non-spherical perturbations. Here we consider the nonlinear case in
{\em Eulerian} coordinates. 

To answer both questions we use the mathematical methods of a classic
paper by Carrier and Greenspan \cite{CarrierGreenspan} concerning the
shallow water equations on a sloping beach. We begin by reviewing
their results and extending them from the shallow water case $n=1$ to
the general polytropic case $n>0$.

%%%%%%%%%%%%%%%%%%%%%%%%%%%%%%%%%%%%%%%%%%%%%%%%%%%%%%%%%%%%%

\section{Mathematical setup}

%%%%%%%%%%%%%%%%%%%%%%%%%%%%%%%%%%%%%%%%%%%%%%%%%%%%%%%%%%%%%

For simplicity we assume spherical symmetry. Near the surface of the
star, gravity is typically weak. Furthermore, the formation of shocks
does not require large fluid velocities. This suggests that Newtonian
physics should be a good approximation for what we want to
investigate. On a sufficiently small scale the spherical symmetry of
the star reduces to planar symmetry, and the Newtonian gravitational
acceleration $g$ is dominated by the interior of the star, and can be
approximated as constant in space and time. Finally, for smooth
solutions, sufficiently close to the surface, the entropy gradient can
be neglected compared to the density gradient in determining the
pressure gradient. We can therefore approximate the ideal gas as
isentropic, with equation of state (\ref{polytrope}). (This last
approximation would not hold if a shock reached the surface.)

In the ``radial'' spatial coordinate $x$ and time $t$, with $v$ the
Eulerian fluid velocity in the $x$ direction, the Euler and
conservation equations are
\begin{eqnarray}
v_t+vv_x+\Gamma K\rho^{1/n-1}\rho_x&=&-g, \\
\rho_t+v\rho_x+\rho v_x &=&0.
\end{eqnarray}
Here $\rho=0$ defines a free boundary $x=x_*(t)$. Within the approximation
of planar symmetry, $x$ has an infinite range, with $x<x_*(t)$
representing the interior of the star. 

It is useful to replace the dependent variable $\rho$ with the sound
speed $c$ given by $c^2=dP/d\rho=\Gamma K\rho^{1/n}$ to obtain
\begin{eqnarray}
\label{vdot}
v_t+vv_x+2ncc_x&=&-g, \\ 
\label{cdot}
c_t+vc_x+{1\over 2n}cv_x&=&0.  
\end{eqnarray}

For $n=1$, these equations are identical with the shallow water
equations restricted to planar symmetry on a uniformly sloping beach,
with $x$ and $v$ the horizontal position and velocity, $\rho\sim c^2$
the height of the water, $g$ the effective horizontal gravitational
acceleration, and $x=x_*(t)$ the instantaneous shoreline
\cite{Acheson}.

The unique static solution of (\ref{vdot},\ref{cdot}) is
\begin{equation}
\label{static}
v=0, \quad c=\sqrt{-{gx\over n}},
\end{equation}
and hence $\rho\sim(-x)^n$, where we have fixed a translation
invariance by locating the surface at $x=0$.

%%%%%%%%%%%%%%%%%%%%%%%%%%%%%%%%%%%%%%%%%%%%%%%%%%%%%%%%%%%%%

\section{Hodograph transform}

%%%%%%%%%%%%%%%%%%%%%%%%%%%%%%%%%%%%%%%%%%%%%%%%%%%%%%%%%%%%%

The problem can be written as
\begin{equation}
\label{Riemann}
\left[\p_t+(v\pm c)\p_x\right](v+gt\pm 2nc)=0,
\end{equation}
and so admits the Riemann invariants $(v+gt)\pm 2nc$ with
characteristic speeds $v\pm c$. Carrier and Greenspan
\cite{CarrierGreenspan} (considering the shallow water case $n=1$)
suggested a hodograph transform from independent variables $t$ and $x$
to independent variables $\lambda$ and $\sigma$ given by
\begin{eqnarray}
\label{lambdadef}
\l &\equiv& v+gt, \\
\label{sigmadef}
\s &\equiv& 2nc.
\end{eqnarray}
(These definitions differ from \cite{CarrierGreenspan} by a factor of
$2$.)

The resulting transformation of partial derivatives is
\begin{equation}
\left(\begin{array}{c} \p_t \\ \p_x \end{array}\right)
 =\Delta^{-1}\left(\begin{array}{cc} x_\s  & - x_\l  \\ - t_\s  &
  t_\l  \end{array}\right)
\left(\begin{array}{c} \p_\l  \\ \p_\s  \end{array}\right),
\end{equation}
where 
\begin{equation}
\label{Deltadef}
\Delta\equiv t_\l  x_\s -x_\l  t_\s .
\end{equation}
In particular, we have 
\begin{equation}
\label{vartrans}
\left(\begin{array}{cc} \l_t  & \s_t  \\ \l_x  &
  \s_x  \end{array}\right)
=\Delta^{-1}\left(\begin{array}{cc} x_\s  & - x_\l  \\ - t_\s  &
  t_\l  \end{array}\right).
\end{equation}
Clearly, the transformation is regular if and only if $\Delta\ne 0,\pm\infty$.

Substituting (\ref{lambdadef}-\ref{sigmadef}) and (\ref{vartrans})
into (\ref{Riemann}), we obtain
\begin{eqnarray}
\label{xtpde1}
x_\s-(\lambda-gt)\, t_\s+\left({\sigma\over 2n}\right) t_\l &=& 0, \\
\label{xtpde2}
x_\l+\left({\sigma\over 2n}\right) t_\s-(\lambda-gt)\,t_\l &=& 0.
\end{eqnarray}
This PDE system is not yet linear because of the appearance of $gt$ in
the coefficients of $t_\s$ and $t_\l$. However, from the two nonlinear
first-order PDEs (\ref{xtpde1}-\ref{xtpde2}) one can derive a linear
second-order PDE for $t(\lambda,\sigma)$, namely
\begin{equation}
t_{\l\l}=t_{\s\s}+{2n+1\over \s} t_\s.
\end{equation}
Trivially, $\lambda$ taken as a function of $\lambda$ and $\sigma$
obeys the same PDE as $t$, and by adding the two we obtain the
autonomous linear wave equation
\begin{equation}
\label{vwave}
v_{\l\l}=v_{\s\s}+{2n+1\over \s} v_\s
\end{equation}
for $v(\lambda,\sigma)$. This is the key equation of this paper. 

The problem has now been cast into linear form, and the free boundary
$x=x_*(t)$ has been mapped to the coordinate line $\sigma=0$, with
$\sigma>0$ representing the interior of the star.

%%%%%%%%%%%%%%%%%%%%%%%%%%%%%%%%%%%%%%%%%%%%%%%%%%%%%%%%%%%%%

\section{A criterion for shock formation}

%%%%%%%%%%%%%%%%%%%%%%%%%%%%%%%%%%%%%%%%%%%%%%%%%%%%%%%%%%%%%

From (\ref{vartrans}) with (\ref{lambdadef},\ref{sigmadef}) we find
\begin{eqnarray}
\label{vxDelta}
v_x&=&{1\over g\Delta} v_\s, \\
\label{cxDelta}
c_x&=&{1\over 2ng\Delta} \left(1-v_\l\right),
\end{eqnarray}
and so a shock forms from regular initial data as and only if
$\Delta\to 0$.  Using
(\ref{lambdadef}-\ref{sigmadef},\ref{xtpde1}-\ref{xtpde2}), the
Jacobian $\Delta$ defined by (\ref{Deltadef}) can be expressed in
terms of $v$ alone as
\begin{equation}
\label{Deltaval}
\Delta = -{\s\over 2ng^2}\left[\left(1-v_\l\right)^2-v_\s^2\right].
\end{equation}
We see that the wave does not form a shock if the first derivatives of
$v$ in a solution of (\ref{vwave}) remain sufficiently small, so that
$\Delta$ remains negative. Such solutions are easily obtained by
rescaling the amplitude of any given solution.

We shall now consider small smooth initial data for (\ref{vdot},\ref{cdot})
on the curve $t=0$, $x<0$. These correspond to Cauchy data for (\ref{vwave}) on
the curve given by $\lambda=\lambda_0(\sigma)$, $\sigma>0$. We require these data
to obey
\begin{equation}
\label{criterion}
\left(1-v_\l\right)^2-v_\s^2>0
\end{equation}
for all $\rho>0$ on $\lambda=\lambda_0(\sigma)$. This criterion is
necessary for the existence of the equivalence between
(\ref{vdot},\ref{cdot}) and (\ref{vwave}), and implies that there is
no shock present in the initial data. We then formally evolve the data
to $\lambda>\lambda_0(\sigma)$ using (\ref{vwave}). Setting aside the
boundary at $\sigma=0$, which we consider later, this solution exists
because (\ref{vwave}) is linear.  However, if at any point in
$\lambda>\lambda_0(\sigma)$ the condition (\ref{criterion}) is violated,
the wave has developed a shock at some $t>0$, and the solution of
(\ref{vwave}) does not have physical meaning for larger values of $t$.

In order to translate initial data in coordinates $(x,t)$ to
$(\sigma,\lambda)$, we consider smooth data with compact support away
from the boundary and which are sufficiently weak (in the sense of
close to the static star solution) that initially the solution can be
approximated by a solution of the linearisation of
(\ref{vdot},\ref{cdot}) around the static star solution. We then
evolve these data using (\ref{vwave}), and so do not require them to
remain small. We use (\ref{criterion}) in this solution as the
necessary and sufficient criterion for the absence of shocks.

Linearising (\ref{vdot},\ref{cdot}) about the static solution
(\ref{static}), we obtain 
\begin{equation}
\delta v_{tt}=\left(-{gx\over n}\right)\left(\delta v_{xx}+{n+1\over
  x} \delta v_x\right).
\end{equation}
(We have written $\delta v$ instead of $v$ to stress that this
is only an approximation valid for small $v$.) The same equation can
be obtained from (\ref{vwave}) by the substitutions
\begin{equation}
\label{lstx}
\l=gt, \quad \s=2\sqrt{-gnx}.
\end{equation}
This gives us a simple approximate relation between initial data 
for the {\it linearisation} of (\ref{vdot},\ref{cdot}), and initial
data for (\ref{vwave}) (which is linear but contains the nonlinear
dynamics). 

A formal d'Alembert solution of (\ref{vwave}) is
h\begin{equation}
v(\l,\s)=\sum_\pm \sum_{k=0}^\infty\s^{-n-{1\over 2}-k}f_k^\pm(\l\pm\s)
\end{equation}
where $f_0^\pm$ is free data and
\begin{equation}
f_{k+1}^\pm={\left(k+{1\over2}\right)^2-n^2\over 2(k+1)} \int f_k^\pm.
\end{equation}

A few remarks will put this result into context: This series converges
at most in the sense of an asymptotic series as $\s\to\infty$, and
clearly diverges for sufficiently small $\s$. Another formal
d'Alembert solution exists which has {\em ascending} powers of $\s$,
but it does not interest us here. In the special case $n=1/2$, either
series reduces to the well-known d'Alembert solution of the spherical
wave equation in 3 dimensions, while for $n=-1/2$ we obtain the
d'Alembert solution of the 1-dimensional wave equation.

Consider now an isolated wave packet approaching the surface with
initial position $\s_0$, width $\s_1\ll \s_0$ and amplitude $v_0$, so
that $|v_\l|\sim |v_\s| \sim v_0/\s_1$ initially. In this regime, we
can approximate
\begin{equation}
\label{leadingorder}
v(\l,\s)\simeq \s^{-n-{1\over 2}}f_0^+(\l+\s).
\end{equation}
The derivatives of $v$ take their largest values when the wave packet
turns around close to the surface. From the scaling properties of solutions of
(\ref{vwave}), this must happen at $\s\sim\s_1$, at which point its
amplitude will be $v_0 (\s_1/\s_0)^{-n-1/2}$ in the approximation
(\ref{leadingorder}). Evaluating (\ref{criterion}) at that point, we
obtain a criterion for the wave never to form a shock, which is
\begin{equation}
{v_0\over\s_1}\lesssim \left(\s_1\over\s_0\right)^{n+{1\over 2}}.
\end{equation}

Finally, expressing $\s_0$ and $\s_1$ in terms of the initial Eulerian
position $x_0$ and length scale $x_1$ of the wave packet by using
(\ref{lstx}), we obtain the regularity criterion
\begin{equation}
\label{final}
{v_0\over\sqrt{g x_0}}\lesssim \left(x_1\over |x_0|\right)^{n+(3/2)}
\end{equation}
In these estimates we neglect an unknown $O(1)$ factor depending on the
precise shape of the wave packet.

Although we have worked in the approximation of planar symmetry and
constant $g$, it is useful to express the parameter $g$ in terms of
$v_*=\sqrt {2gr_*}$, which is the escape velocity at the surface of a
spherical star, where $r_*$ is its radius and $g$ is the gravitational
acceleration at its surface. We can then rewrite the estimate
(\ref{final}) as
\begin{equation}
v_0\lesssim \left(x_1\over |x_0|\right)^{n+(3/2)} \left(|x_0|\over
r_*\right)^{1\over 2}v_*
\end{equation}

A numerical example will illustrate this: in a neutron star modelled as
a polytrope with $r_*\sim 10^4m$, $v_*\sim 10^8m/s$ and $n=1$, a sound
wave of wavelength $x_1\sim 1m$ deep in the interior ($x_0\sim -r_*$)
must have an amplitude of $v_0\lesssim 10^{-2} m/s$ to remain regular. 

%%%%%%%%%%%%%%%%%%%%%%%%%%%%%%%%%%%%%%%%%%%%%%%%%%%%%%%%%%%%%

\section{Generic behaviour at the free boundary}

%%%%%%%%%%%%%%%%%%%%%%%%%%%%%%%%%%%%%%%%%%%%%%%%%%%%%%%%%%%%%

The surface of the star is a free boundary characterised by the
kinematic boundary conditions
\begin{eqnarray}
P(x_*(t),t)&=&0, \\ 
{dx_*\over dt}&=&v(x_*(t),t).
\end{eqnarray}
These conditions are straightforward to implement in Lagrangian
coordinates, but in 3D HRSC simulations we need their equivalent in
Eulerian coordinates. For solutions which remain smooth, we obtain
these by going through the hodograph transformation.

The general solution of Eq.~(\ref{vwave}) can be written as a linear
superposition of solutions of the form
\begin{equation}
\label{Besselsolution}
v(\l,\s)=e^{i\omega\l}\,\s^{-n} J_{\pm n} (\omega\s).
\end{equation}
As $J_n(\sigma)$ is $\s^n$ times a power series in positive even
powers of $\sigma$, the solution using $J_n$ is an even regular
function of $\s$, while the solution using $J_{-n}$ diverges as
$\s^{-2n}$ as $\s\to 0$. The regular solution can be selected by
imposing the boundary condition
\begin{equation}
\label{BC}
v_\s=0 \quad \hbox{at} \quad \s=0,
\end{equation}
which together with (\ref{vwave}) makes a well-posed linear
initial-boundary value problem. Clearly this condition is the required
kinematic boundary condition for smooth solutions. 

We now translate this back into the Eulerian variables $c(x,t)$ and
$v(x,t)$.  Assuming the wave does not form a shock, the square bracket
in (\ref{Deltaval}) is strictly positive, and so $\Delta\sim\s$ at the
boundary. Substituting $\Delta\sim\s$ into (\ref{cxDelta}) gives
\begin{equation}
\s c_x\sim \left(1-v_\l\right)
\end{equation}
at the boundary. The right-hand side is even in $\s$ because $v$ is
even in $\s$ by the assumption of regularity. It follows, using
(\ref{sigmadef}),  that
$(c^2)_x$ is a regular function of $c^2$, and hence $c^2$ is a regular
function of $x$.

Substituting $\Delta\sim\s$ into 
(\ref{vxDelta}) gives
\begin{equation}
v_x\sim \s^{-1}v_\s 
\end{equation}
at the boundary. The right-hand side is again even in $\s$. Hence
$v_x$ is an even function of $c^2$ and so, using our previous result,
it is a regular function of $x$. It follows that $v$ is a regular
function of $x$.

It is clear that the $\l$ or $t$ dependence does not affect these
results in the limit $x\to x_*(t)$ or $\sigma\to 0$. We have therefore
shown that as long as the solution remains regular, $c^2$ and $v$ are
regular functions of $x$ and $t$ at the surface. This is the desired
kinematic free boundary condition. In particular, $c^2\sim x_*(t)-x$
at the moving surface of regular solutions, as in the static case.

Note that $\rho\sim (c^2)^n$, so $\rho$ is a regular function of $x$ only
if $n$ is an integer. Note also that in general $v$ and $c^2$ are neither even
nor odd in $x-x_*$. 

%%%%%%%%%%%%%%%%%%%%%%%%%%%%%%%%%%%%%%%%%%%%%%%%%%%%%%%%%%%%%%%%%%%%%%%%%%%%

\section{Discussion}

%%%%%%%%%%%%%%%%%%%%%%%%%%%%%%%%%%%%%%%%%%%%%%%%%%%%%%%%%%%%%%%%%%%%%%%%%%%%

Building on the earlier work \cite{CarrierGreenspan,PP}, we have given
various forms of an upper limit on the amplitude of nonlinear sound
waves if they are to avoid forming a shock. This tells us in which
physical regime a simple (non-shock capturing) numerical method will
be valid because shocks do not occur. It may also be of direct
astrophysical interest. 

For solutions which remain regular as they are reflected
at the free boundary, we have shown that the usual free boundary
condition is equivalent to $v$ and $c^2$ being regular functions of
$x$ and $t$. This suggests an alternative numerical
treatment of the stellar surface which does not require an unphysical
atmosphere.

Our results were derived within the approximations of Newtonian
physics, a constant gravitational field, a polytropic equation of
state and planar symmetry (as the limit of spherical symmetry near the
surface). As discussed in the introduction, these are all natural
approximations to make, except for spherical symmetry. However,
applying geometric optics to the linearised sound wave equation for
the pressure perturbation $\delta P$,
\begin{equation}
\delta P_{tt}=\left(-{gx\over n}\right) \left(\delta P_{xx}+{n+1\over
  x} \delta P_x+\delta P_{yy}+\delta P_{zz}\right), 
\end{equation}
we find that its sound rays, without loss of generality restricted to
the $xy$ plane, are given by $y(x)=a+\ln(1+bx^2)$ for constants $a$
and $b$, and so in the geometric optics approximation sound waves
approaching the surface $x=0$ at any angle are refracted towards lower
sound speed until they reach the surface at right angles. This
provides some justification for the assumption that our results will
also be qualitatively correct beyond the restriction to spherical
(planar) symmetry.

%%%%%%%%%%%%%%%%%%%%%%%%%%%%%%%%%%%%%%%%%%%%%%%%%%%%%%%%%%%%%%%%%%%%%%%%%%%%

\acknowledgments

We would like to thank Marvin Jones for discussions, and Michael Gabler for
pointing out an error in the original version. 

%%%%%%%%%%%%%%%%%%%%%%%%%%%%%%%%%%%%%%%%%%%%%%%%%%%%%%%%%%%%%

%%%%%%%%%%%%%%%%%%%%%%%%%%%%%%%%%%%%%%%%%%%%%%%%%%%%%%%%%%%%%

\end{document}